\documentclass[journal]{IEEEtran}
\usepackage{multirow} 
\usepackage{xcolor,soul,framed} 
\usepackage[linesnumbered,ruled,vlined]{algorithm2e} 
\colorlet{shadecolor}{yellow}

\usepackage[pdftex]{graphicx}
\graphicspath{{../pdf/}{../jpeg/}}
\DeclareGraphicsExtensions{.pdf,.jpeg,.png}

\usepackage[cmex10]{amsmath}
\usepackage{array}
\usepackage{mdwmath}
\usepackage{mdwtab}
\usepackage{eqparbox}
\usepackage{url}
\usepackage{comment}

\begin{document}
\bstctlcite{IEEEexample:BSTcontrol}
    \title{Influence Maximization in Hypergraphs Using A Genetic Algorithm with New Initialization and Evaluation Methods}
    
\author{
    Xilong Qu, 
    Wenbin Pei, 
    Yingchao Yang, 
    Xirong Xu, 
    Renquan Zhang, and 
    Qiang Zhang
    \vspace{-0.5cm}
   	\thanks{Xilong Qu, Wenbin Pei, Yingchao Yang, Xilong Xu and Qiang Zhang are with the School of Computer Science and Technology, Dalian University of Technology, Dalian 116024, China; Key Laboratory of Social Computing and Cognitive Intelligence (Dalian University of Technology), Ministry of Education, Dalian 116024, China (e-mail: quxilong@mail.dlut.edu.cn; peiwenbin@dlut.edu.cn; nmmprepeat@gmail.com; xirongxu@dlut.edu.cn; zhangq@dlut.edu.cn). }

    \thanks{Renquan Zhang is with the School of Mathematical Sciences, Dalian University of Technology, Dalian 116024, China (e-mail: zhangrenquan@dlut.edu.cn). 
    
    This work was supported in part by the National Key Research and Development Program of China under grant 2021ZD0112400, the National Natural Science Foundation of China under grants 62206041, 12371516 and U21A20491, and the NSFC-Liaoning Province United Foundation under grant U1908214, the 111 Project under grant D23006, the Liaoning Revitalization Talents Program under grant XLYC2008017, and China University Industry-University-Research Innovation Fund under grants 2022IT174, Natural Science Foundation of Liaoning Province under grant 2023-BSBA-030, and an Open Fund of National Engineering Laboratory for Big Data System Computing Technology (Grant No. SZU-BDSC-OF2024-09).
}
		 
}    

\markboth{JOURNAL OF LATEX CLASS FILES}{Roberg \MakeLowercase{\textit{Qu et al.}}:Influence Maximization in Hypergraphs}

\maketitle

\begin{abstract}
Influence maximization (IM) is a crucial optimization task related to analyzing complex networks in the real world, such as social networks, disease propagation networks, and marketing networks. Publications to date about the IM problem focus mainly on graphs, which fail to capture high-order interaction relationships from the real world. Therefore, the use of hypergraphs for addressing the IM problem has been receiving increasing attention. However, identifying the most influential nodes in hypergraphs remains challenging, mainly because nodes and hyperedges are often strongly coupled and correlated. In this paper, to effectively identify the most influential nodes, we first propose a novel hypergraph-independent cascade model that integrates the influences of both node and hyperedge failures. Afterward, we introduce genetic algorithms (GA) to identify the most influential nodes that leverage hypergraph collective influences. In the GA-based method, the hypergraph collective influence is effectively used to initialize the population, thereby enhancing the quality of initial candidate solutions. The designed fitness function considers the joint influences of both nodes and hyperedges. This ensures the optimal set of nodes with the best influence on both nodes and hyperedges to be evaluated accurately. Moreover, a new mutation operator is designed by introducing factors, i.e., the collective influence and overlapping effects of nodes in hypergraphs, to breed high-quality offspring. In the experiments, several simulations on both synthetic and real hypergraphs have been conducted, and the results demonstrate that the proposed method outperforms the compared methods.
\end{abstract}


%
\IEEEpeerreviewmaketitle


\section{Introduction}

\IEEEPARstart{T}{he} complex network theory has been extensively studied due mainly to its widespread applications in diverse domains, including social networks \cite{Boyd2007,Lee2014,Cai2022}, energy networks \cite{Blaabjerg2006,Jiang2022,Olmi2024,Grzybowski2018}, and epidemic transmission networks \cite{Li2020,Pei2021,Tai2023,Chaharborj2022}. However, due to the complexity and heterogeneity of a network structure \cite{Brockmann2013}, only a few key nodes play a vital role in the propagation process. 
For example, in social networks \cite{Al2017}, influential individuals typically possess professional expertise, significant persuasive abilities, and extensive social connections. Their statements and behaviors rapidly disseminate throughout the network, eliciting widespread attention and discussions, thus exerting a profound influence on social opinions and individual behaviors. Identifying such influential nodes poses a fundamental problem known as Influence Maximization (IM).

The IM problem involves selecting $k$ nodes to maximize propagation based on a set of propagation rules. It is a typically discrete optimization problem and has been proven to be NP-hard \cite{Domingos2001}. 
Approaches to the IM problem are broadly classified into four main categories, including greedy-based algorithms \cite{Goyal2011,Lozano-Osorio2023}, centrality-based algorithms \cite{Kundu2011}, community-based algorithms \cite{Yu2010,Kumar2022}, and meta-heuristic algorithms \cite{Gong2016-1,Wang2021}. Greedy-based algorithms have demonstrated effectiveness in tackling the IM problem. Nevertheless, their high computational complexity restricts their applicability to large-scale networks. Community-based and meta-heuristic algorithms have been proposed to balance a trade-off between accuracy and time complexity. However, due to the complexity of network structures and propagation dynamics, identifying the most influential nodes remains a challenging task. 

Publications to date on the IM problem focus primarily on graphs. As advancements in modeling complex networks have progressed \cite{Torres2021}, there has been a recognition that conventional graphs are limited to representing only simple binary relationships. This poses a new open challenge in modeling higher-order interactions \cite{Battiston20201}.
For example, in social networks \cite{Firouzkouhi2024}, when interpersonal connections extend beyond simple links to encompass intricate dynamics such as group affiliations and shared interests, conventional graphs fall short in capturing the entirety of these complexities. To overcome this limitation, researchers have introduced hypergraphs. Unlike traditional graphs, hypergraphs employ hyperedges to represent higher-order relationships, connecting multiple nodes simultaneously. This enables more effective modeling of intricate interactions, as each hyperedge encapsulates a group of nodes participating in a collective interaction or event. 
However, identifying the most influential nodes in hypergraphs has been less investigated but remains challenging, mainly because nodes and
hyperedges are often strongly coupled and correlated. Furthermore, it is also challenging to characterize the intricate interactions between hyperedges and nodes for developping an efficient algorithm to tackle IM problems.

Genetic algorithms (GAs) search for optimal solutions to problems by simulating mechanisms such as natural selection and biological evolution. GAs have effectively addressed complex optimization problems, showcasing robust global search capabilities \cite{Grefenstette1986,Wang2010,Pizzuti2012418}. This inspires us to use GAs to identify nodes that exert optimal influence on both nodes and hyperedges in the IM problem. 

In this paper, we proposed a new method, named \textbf{G}enetic Algorithm with \textbf{C}ollective \textbf{I}nfluence \textbf{I}nitialization and Integrated \textbf{M}utation Operation (G-CIIM), to tackle the IM problem in hypergraphs. The major contributions of this paper are outlined as follows:

\begin{enumerate}

\item We present a new Independent Cascade (IC) Model to hypergraphs for the first time, simultaneously accounting for the impact of both nodes and hyperedges in the propagation of cascading failures. Because cascading effects from the hyperedges and nodes are integrated, the IC model offers a more comprehensive understanding of influence propagation dynamics on hypergraph structures. The proposed G-CIIM method is based on the framework structured by the IC model for discerning between influential nodes in hypergraphs. It is noticed that the proposed IC model can provide a general model framework for IM problems to identify influential nodes in hypergraphs, not limited to G-CIIM only.

\item We design a new initialization method for G-CIIM, which leverages the hypergraph collective influence to initialize a population of good-quality initial solutions at the beginning of the evolutionary learning process. Experimental simulations show that the initialization method can accelerate the convergence speed of the algorithm.

\item We design a new fitness function, which concurrently considers the joint influences of both nodes and hyperedges, to assess the influence of the seed set. This fitness function facilitates the selection of superior individuals within a population. Experimental results indicate that the new fitness function enables G-CIIM to effectively identify the most influential nodes for both nodes and hyperedges.

\item We design a mutation operator that considers the aspects of collective influence and overlapping effects of nodes, thereby effectively traversing the solution space. Empirical results on synthetic hypergraphs demonstrate that G-CIIM can yield superior solutions.

\end{enumerate} 

The remainder of this paper is structured as follows. Section II provides a comprehensive review of related works in the field. Then, we present our proposed G-CIIM method in more detail in Section III. In Section IV, we verify the effectiveness of G-CIIM on synthetic and real-world hypergraphs, respectively. Finally, Section V concludes the major contributions of our study, and discusses implications and potential avenues for future research.

\section{Background}
\subsection{IM Problems}

In IM problem, a complex network, e.g., social networks, energy networks, and citation networks, can be represented by a graph $G(V, E)$, where $V$ signifies a set of nodes and $E$ signifies a set of edges. Given a positive integer $k$ that is much smaller than the total number of nodes $(N=|V|)$, the objective is to select $k$ nodes from the network as a seed set $S$ for activating the maximum number of nodes under a fixed propagation rule. The IM problem can be mathematically defined as follows:

\begin{equation}
\begin{array}{l}
S^* = \arg \max (\sigma (S))\\
st.\quad \left| S \right| = k\\
\quad \quad S \subseteq V
\end{array}
\label{IM-DF}
\end{equation}
where $\sigma(S)$ denotes the number of activated nodes when the seed set $S$ is used for influence propagation, and $S^*$ is the influence maximization seed set.

The propagation models serve as the foundation for the study of IM problems. The linear threshold (LT) and IC models proposed in \cite{Kempe2003} are the two most classical models for studying IM problems in graphs. The network structure plays a particularly critical role in the propagation process. Possible network structure methods encompass betweenness centrality \cite{Brandes2001}, closeness centrality \cite{Okamoto2008}, degree centrality \cite{Opsahl2010}, and PageRank centrality \cite{Page1998}. Centrality-based methods attempt to identify influential nodes by evaluating their structural importance within a network. Despite their widespread use, these methods frequently perform less precisely. To overcome this limitation, Kempe et al. \cite{Kempe2003} proposed a greedy algorithm that iteratively selects nodes to maximize influence spread, yielding high-precision solutions. However, the greedy algorithm is computationally intensive, particularly for large-scale networks. To reduce computational costs, several variant algorithms have been designed, including CELF++ \cite{Goyal2011}, the SMG algorithm \cite{Heidari2015}, and the TSIFIM algorithm \cite{Dong2023}. Nonetheless, it is still challenging to develop an algorithm that effectively balances the accuracy and efficiency for addressing the IM problems.

Evolutionary algorithms (EAs) search for optimal solutions through simulating biological processes such as natural selection, crossover, and mutation. There have been EA-based algorithms \cite{Gong2016,Liu2019,Wang2021,Jiang2011,Gong2016-1} proposed to address the IM problem, to achieve a trade-off between precision and computational efficiency. However, these IM researches focus only on graphs, which overlook higher-order relationships intrinsic to real-world contexts \cite{Li2018}. Hypergraph-based modeling provides a more accurate portrayal of information propagation dynamics among nodes.

\subsection{Existing Hypergraphs-based Works in IM Problems}
\begin{figure}
	\centering
		\includegraphics[scale=.38]{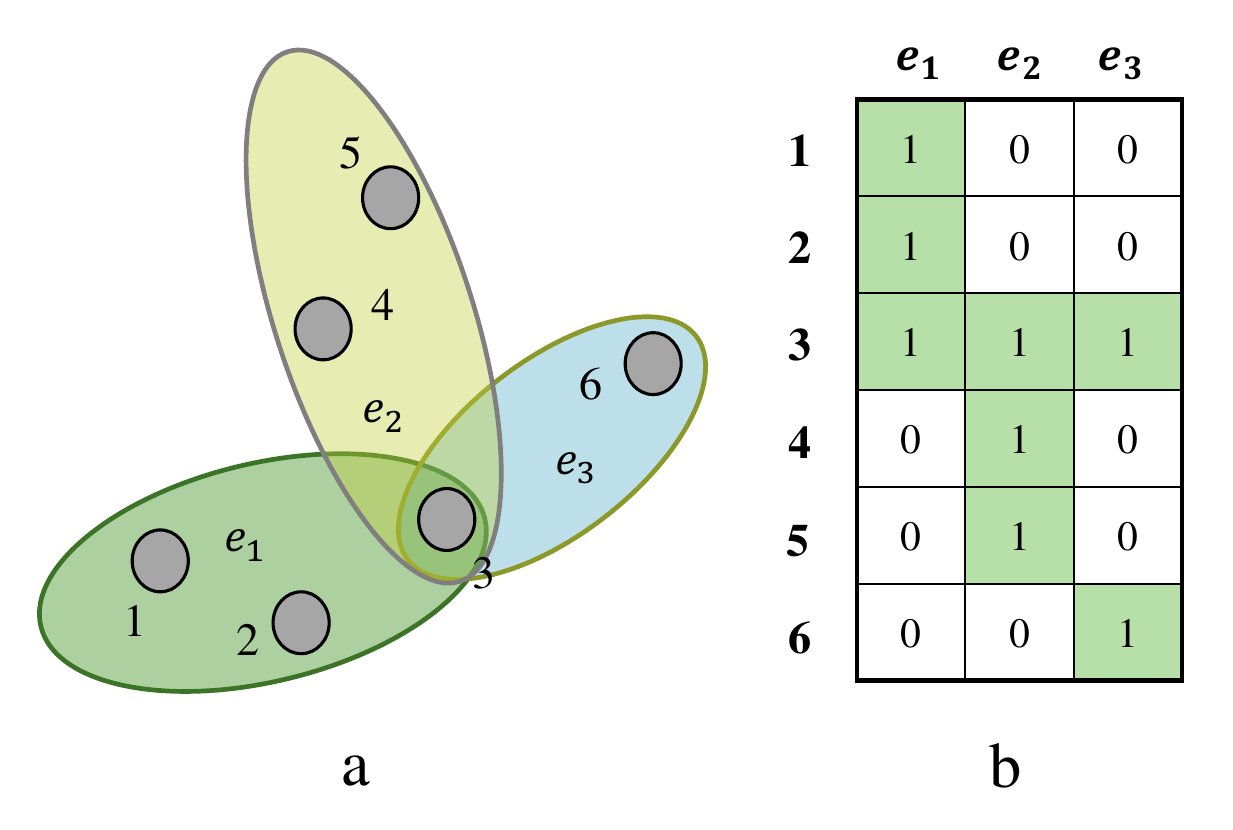}
	\caption{An example of a hypergraph. (a) A hypergraph with $6$ nodes and $3$ hyperedges. (b) The corresponding incidence matrix for the hypergraph.}
	\label{FIG:example}
\end{figure}

A hypergraph with $N$ nodes and $M$ hyperedges can be represented as $H(V, E)$, where $V$ is a set of nodes and $E$ is a set of hyperedges. The number of hyperedges associated with node $i$, denoted as $k_i$, refers to the hyperdegree of node $i$, and it follows a distribution $P(k)$. The number of nodes contained in a hyperedge $e_\gamma$, denoted as $m_\gamma$, is the cardinality of the hyperedge, and it follows a distribution $R(m)$. A hypergraph can be represented by an incidence matrix $H=\{H_{i e_\gamma}\}_{N \times M}$, where $H_{i e_\gamma} = 1$ when there is an associated relationship between node $i$ and hyperedge $e_\gamma$, otherwise $H_{i e_\gamma} = 0$. A hypergraph with $6$ nodes and $3$ hyperedges is illustrated in Fig \ref{FIG:example}.

The IM problem in hypergraphs aims to select $k$ nodes to maximize influence spread under a specified propagation rule. According to the above definition on a hypergraph, it can be found that a hyperedge associates multiple nodes at the same time. This enables the hypergraph representation to distinguish high-order interactions between nodes. Due to this advantage, IM in hypergraphs has been receiving increasing attention. To date, there have been some researches proposed to use hypergraphs for the IM problems. 

Xie et al. \cite{Xie2023} proposed a degree-based adaptive influence maximization algorithm with a low time complexity and good results in synthetic and real-world hypergraphs. However, this research relies on a dynamic simulation in the process of selecting seeds. Puzis et al.\cite{Puzis2013} extended betweenness centrality from graphs to hypergraphs, but its high time complexity prevents it from being applied to large-scale hypergraphs. Zhang et al. \cite{Zhang2024} proposed an analysis framework based on the hypergraph collective influence theory and designed a greedy algorithm to solve the IM problem. This method strikes a balance between efficiency and precision, while it is effective for the hypergraph threshold model only. Although many approaches \cite{Gong2024,Wu2023,Kovalenko2022} have been proposed for IM problems in hypergraphs, they perform less effectively due to the complexity of propagation dynamics and the strong coupling characteristics between nodes and hyperedges.


EA-based methods show a strong global search capability when addressing complex optimization problems \cite{Wang202352}. GA is the most popular EAs, inspired by natural selection and biological evolution. In this study, we employ GAs to address the IM problem in hypergraphs, aiming to identify the most influential node for both nodes and hyperedges.
\begin{figure*}
	\centering
	\includegraphics[scale=.468]{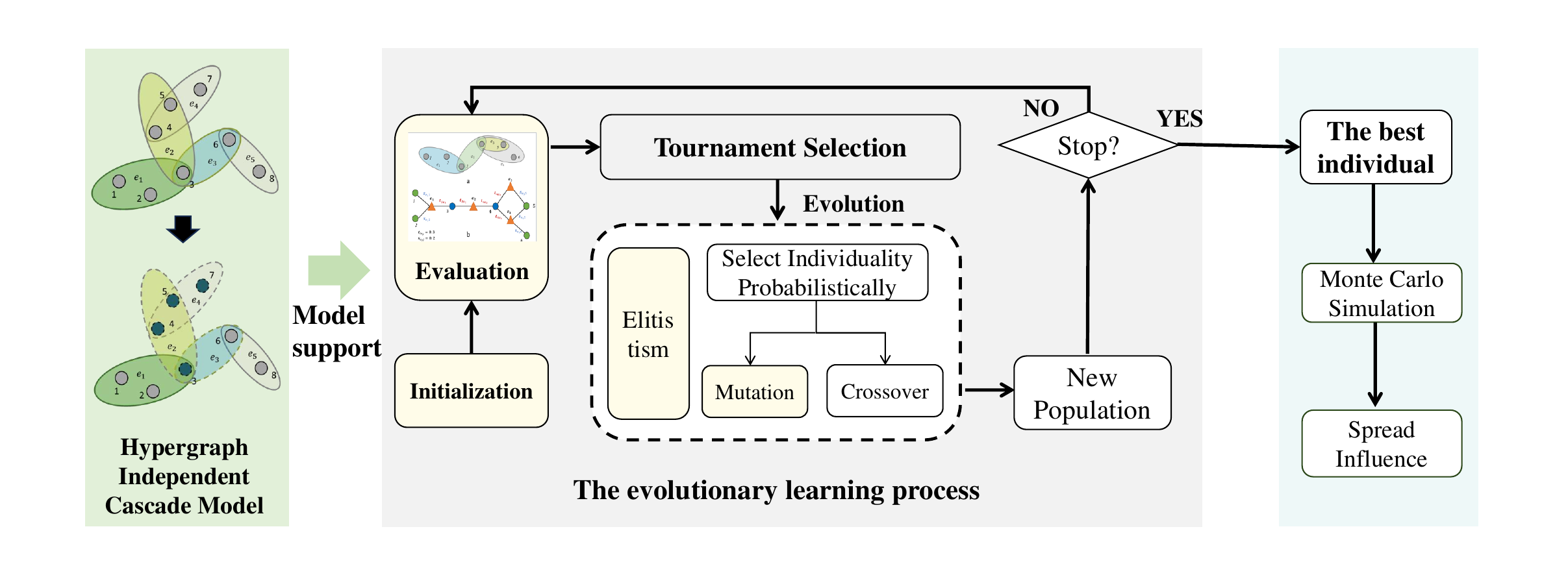}
	\caption{The overall framework diagram of the G-CIIM algorithm.}
	\label{FIG:framework}
\end{figure*}

\section{The Proposed Method}
In this section, we introduce the proposed G-CCIM method. The overall flowchart of the G-CIIM method is shown in Fig \ref{FIG:framework}. Before the evolutionary optimization process, the Hypergraph IC model is designed to describe the information propagation process between nodes and hyperedges. Subsequently, a new initialization method is introduced to G-CIIM, which leverages the hypergraph collective influence to improve the quality of the initial population. To accurately evaluate the goodness of each individual, a new fitness function, which concurrently considers the joint influences of both nodes and hyperedges, is introduced. Then, we introduce a new mutation operator that considers the aspects of collective influence and overlapping effects of nodes.


\subsection{The IC Model in Hypergraphs}
A hypergraph is composed of two basic elements, i.e., nodes and hyperedges, representing different components and the cooperative or competitive relationships between multiple components respectively. In many scenarios, hyperedges are particularly important to allow interactions between multiple nodes \cite{Torres2021,Firouzkouhi2024}. To date, only a few studies \cite{Liu2023} consider the role of both nodes and hyperedges in the propagation process. Moreover, the IC Model is the most classic propagation model for studying the IM problem. However, all the existing studies only investigate IC models in graphs. In the literature, there have not been any studies to investigate how the IC model in graphs can be extended to hypergraphs. 

\begin{figure*}
	\centering
		\includegraphics[scale=.38]{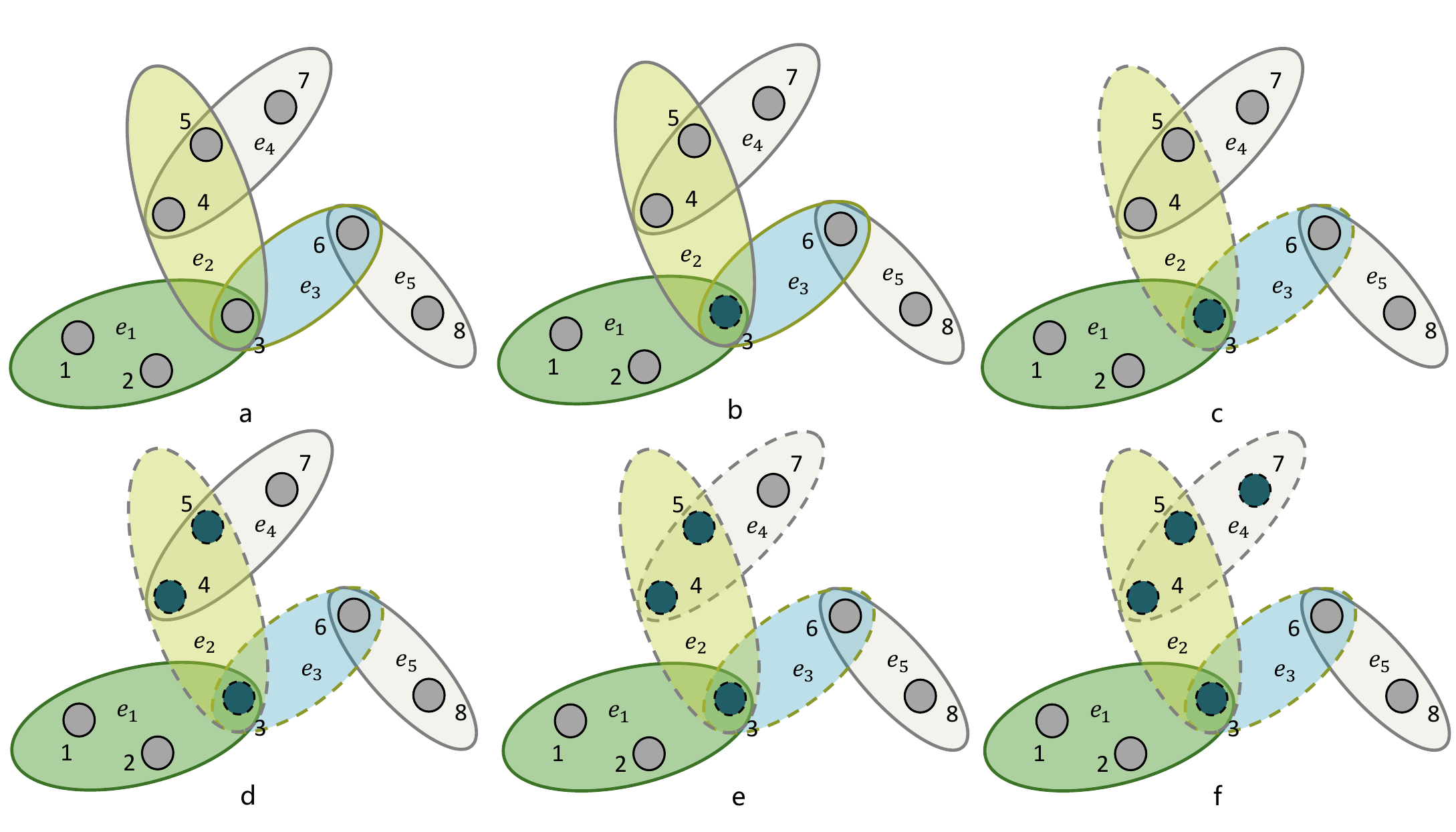}
	\caption{The image depicts the propagation rule of the IC Model in hypergraph. Circles represent nodes, while ovals represent hyperedges. Initially, node $3$ is designated as the seed node, and the propagation probabilities $t_{i e_\gamma}$ and $s_{e_\gamma i}$ are set to $0.5$. As the process begins, the failure of node $3$ leads to the failure of hyperedges $e_2$ and $e_3$. In the subsequent time step, the failure of hyperedge $e_2$ leads to the failure of nodes $4$ and $5$. The failure of node $5$ subsequently causes hyperedge $e_4$ to fail. Finally, the failure of hyperedge $e_4$ leads to the failure of node $7$. At this time, no nodes or hyperedges will fail, and then the propagation process terminates.}
	\label{FIG:1}
\end{figure*}

In hypergraphs, nodes can be treated as components, and hyperedges can be seen as functional modules composed of multiple components. When node (component) $i$ does not work, the associated hyperedge $e_\gamma$ will fail with a probability of $t_{i e_\gamma}$. Conversely, when hyperedge (functional module) $e_\gamma$ fails, its associated node (component) $i$ will fail with a probability of $s_{e_\gamma i}$. Here, $t_{i e_\gamma}$ represents the fault tolerance of hyperedges (functional modules), and $s_{e_\gamma i}$ represents the survivability of nodes (components). At the initial stage, the failures of nodes located in different hyperedges trigger a cascade of fault propagation. This will further lead to the failures of some hyperedges, constituting the propagation process from nodes to hyperedges. Similarly, hyperedges' failures can result in certain nodes' failures, constituting the propagation process from hyperedges to nodes. The propagation process terminates when there are no new failures either from nodes or hyperedges. The propagation rule of the IC Model is illustrated in Fig \ref{FIG:1}.

\subsection{Solution Initialization}

In G-CIIM, the selection information of seed nodes is encoded into a GA individual represented by a vector $S$. The length of an individual is equal to $k$ ($k$ is the number of non-duplicate nodes selected from the total $N$ nodes in a hypergraph). The value of each gene in an individual is the index of a seed node.  


A well-designed initialization strategy can effectively compress the search space, accelerating algorithm convergence. In previous works \cite{Wang2021900,Wang2021,Wang202352}, the degree of nodes has been frequently employed for initializing the population. However, the degree-based algorithm cannot describe the collective influence of multiple nodes. To address this issue, we incorporate the analysis framework from the Message Passing theory into the IC Model within hypergraphs. Then, a collective influence measure with low computational complexity is obtained, utilizing the local structural information of hypergraphs to assess the influence of the node. \textbf{The detail of derivation is introduced in Appendix A}. Through this derivation, the 1-order hypergraph collective influence based on independent cascade model ($HCI-ICM$) of nodes $i$ is defined as follows:

\begin{equation}
HC{I_1}(i) - ICM = {k_i} + \sum\limits_{{e_\gamma } \in \partial i} {{t_{i{e_\gamma }}}({m_\gamma } - 1)}
\label{HCI1-1}
\end{equation}

Where $\partial i$ represents the hyperedges set associated with node $i$. The 2-order hypergraph collective influence of nodes $i$ can be obtained as follows:

\begin{equation}
\begin{aligned}
    HCI_2(i) - ICM &= {k_i} + \sum\limits_{{e_\gamma } \in \partial i} {{t_{i{e_\gamma }}}({m_\gamma } - 1)} \\
    &+ \sum\limits_{{e_\gamma } \in \partial i} {{t_{i{e_\gamma }}}\sum\limits_{j \in {e_\gamma }/i} {(1 - {n_j}){s_{{e_\gamma }j}}({k_j} - 1)} }
\end{aligned}
\label{HCI2-1}
\end{equation}

\begin{figure*}
	\centering
		\includegraphics[scale=.3]{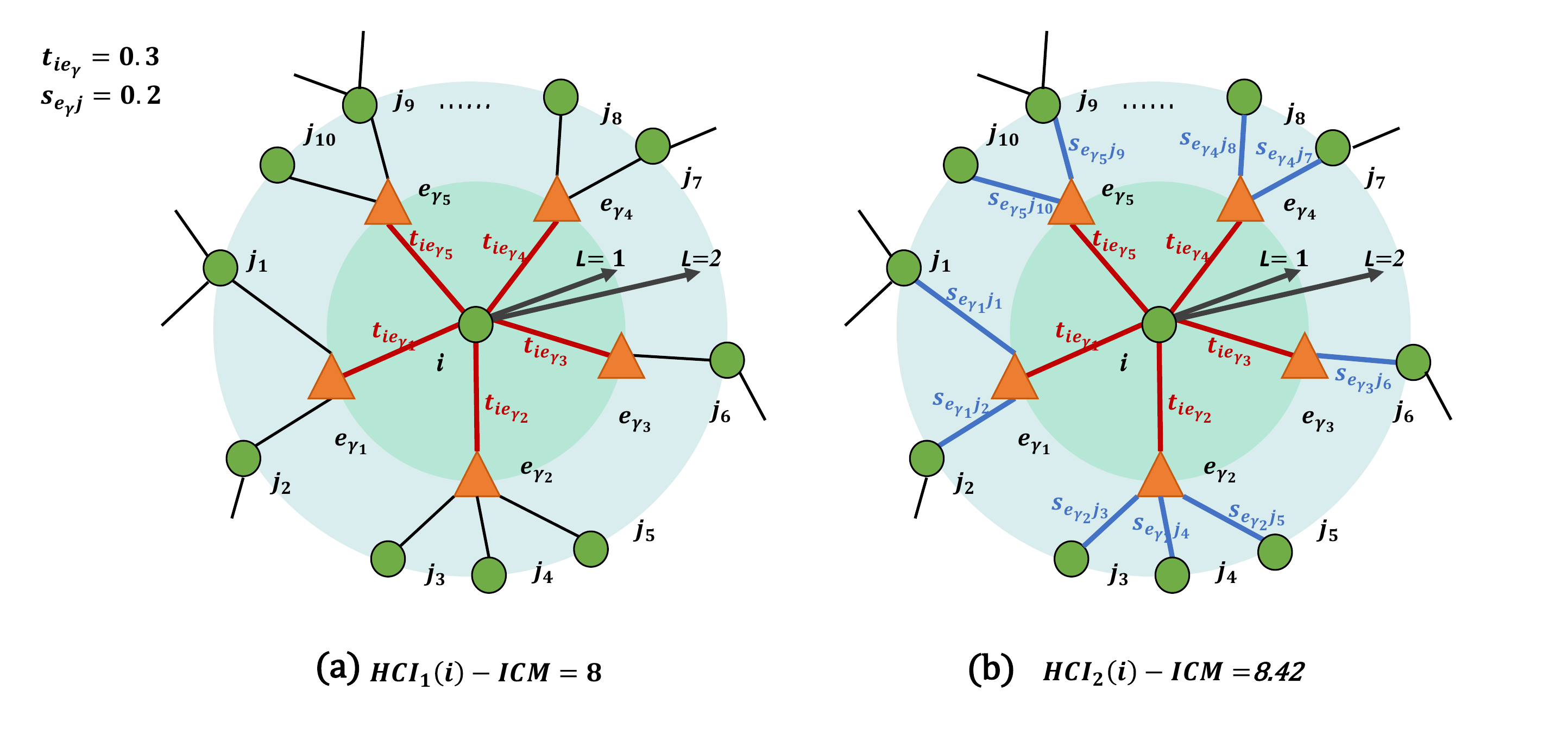}
	\caption{The illustrates of hypergraph collective influence. The green circles represent the nodes, the orange triangles represent the hyperedge, and the edges represent the relationship between the node and the hyperedge. (a) shows the 1-order hypergraph collective influence of node $i$ that can be delineated into two components. The first part, denoted as $k_i$, represents the hyperdegree of node $i$, indicating the number of hyperedges directly connected to node $i$. The second part represents the sum of the probabilities of node $i$ propagating to each associated hyperedge $e_\gamma$, multiplied by the cardinality of the hyperedge $e_\gamma$ minus one. (b) shows the 2-order hypergraph collective influence of node $i$ that also can be defined as the sum of two components. The first component is the 1-order hypergraph collective influence of node $i$. The second is the sum of the probabilities of node $i$ propagating to each associated hyperedge $e_\gamma$, multiplied by the probability of hyperedge $e_\gamma$ propagating to each associated node $j$, and further multiplied by the hyperdegree of node $j$ minus one, summed over all hyperedges $e_\gamma$ and nodes $j$, where $e_\gamma \in \partial i$, $j \in {e_\gamma }/i$. 
}
\label{FIG:3}
\end{figure*}

Where $e_\gamma/j$ represents the set of all the other nodes in the hyperedge $e_\gamma$ except node $j$. In Fig \ref{FIG:3}, we take two examples to illustrate the definition of the $HCI-ICM$ of node $i$.

After considering both the accuracy and time complexity, the proposed method introduces the 1-order hypergraph collective influence to initialize the candidate solutions. 
\begin{figure}
	\centering
		\includegraphics[scale=.46]{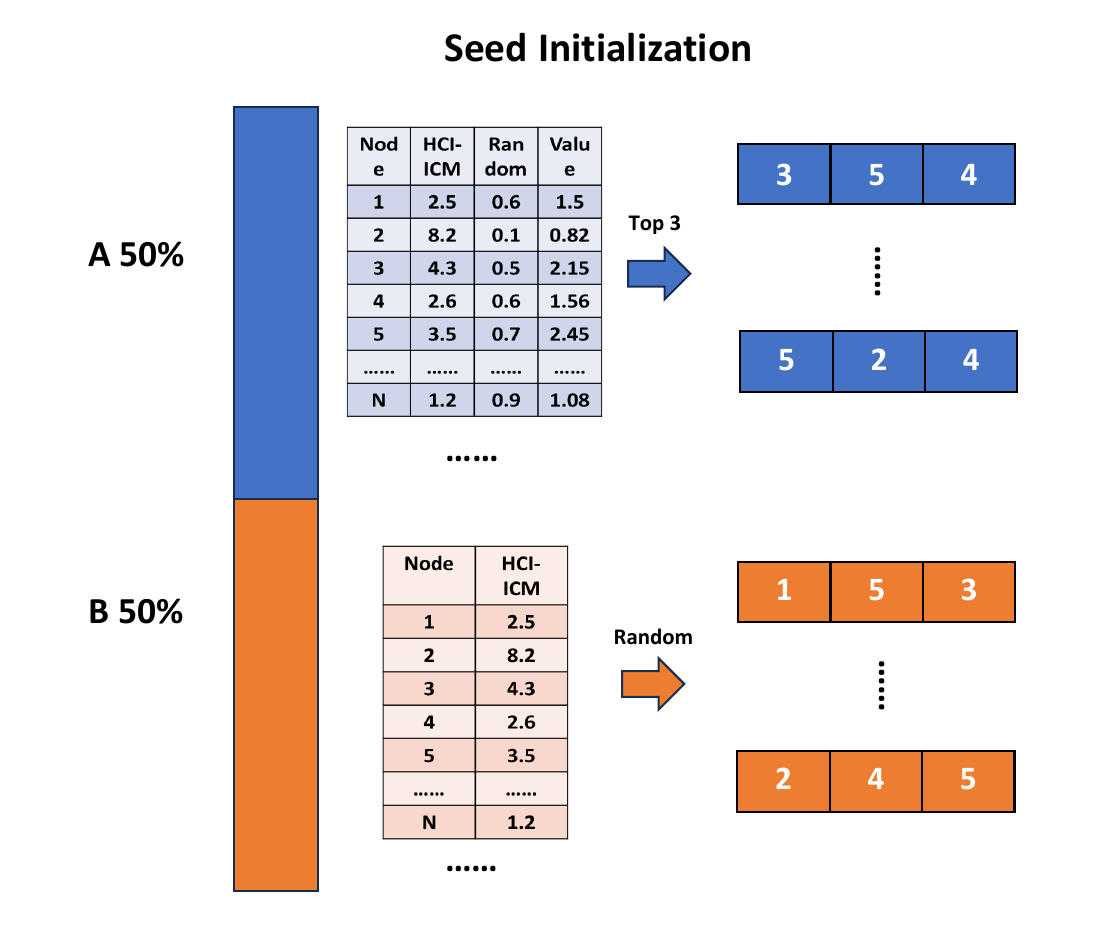}
	\caption{The illustrates of the population initialization process. The population initialization is divided into two parts, part A is that the nodes with large HCI-ICM have more chances to be selected. Part B is the random selection strategy. 
}
\label{FIG:Initialization}
\end{figure}
In the initialization process, two factors are considered. The first is to enhance the quality of the initial population by ensuring that influential key nodes are distributed across different individuals. This thereby speeds up the convergence of the algorithm. The second is to encourage a population's diversity to prevent it from converging prematurely to local optima by introducing randomization. Therefore, the population initialization process has two main stages, as illustrated in Fig. \ref{FIG:Initialization}. In the first stage, the HCI-ICM value of each node is multiplied by a random number between $0$ and $1$. Subsequently, we sort the values and select the top $k$ nodes to develop an individual. This procedure gives influential nodes more chance to be chosen. In the second stage, a random initialization strategy is employed, where $k$ nodes are randomly selected from all the nodes to create an individual. This stage enhances the population diversity, which prevents the population from getting trapped in local optima. The two stages separately generate half of the initial population. 
The Pseudocode of the initialization method is presented in Algorithm \ref{algorithm1}.

\begin{algorithm}
\label{algorithm1}
\SetAlgoLined
\SetKwInOut{Input}{Input}
\SetKwInOut{Output}{Output}
\Input{The size of the seed set IND\_SIZE }
\Output{$Individual$}
\SetKwFunction{FMain}{Initialization}
\SetKwProg{Fn}{Function}{:}{}
\Fn{\FMain{IND\_SIZE}}{
\eIf{random.random() $<$ 0.5}{
    HCI\_array $\leftarrow$ Calculate the 1-order HCI-ICM of all the nodes \;
    HCI\_TMP $\leftarrow$ HCI\_array $\times$ random.random()\;
    $Individual$ $\leftarrow$ Select top IND\_SIZE nodes from HCI\_TMP \;
    \Return $Individual$\;
}{
    Individual $\leftarrow$ Randomly sample IND\_SIZE nodes from all the nodes\;
    \Return $Individual$\;
}
}
\caption{Population Initialization}
\end{algorithm}

\subsection{The Fitness Function}


In IC models, assessing the influence of a set of nodes typically requires Monte Carlo simulations, which are usually computationally expensive. Therefore, it is of great significance to design a low-time complexity fitness function that can approximately estimate the influence of node sets. Gong et al. \cite{Gong2024} show that multi-hop Influence Estimation (MIE) can accurately estimate the influence of nodes. Inspired by MIE and combined with the propagation characteristics of the hypergraph IC model, the fitness function is designed based on the following two factors. The first is to maximize the number of activated nodes, which is the goal of the IM problem. Therefore, the proposed fitness function incorporates the seed set's expected number of activated neighboring nodes. The second is to maximize the seed set's expected number of activated neighboring hyperedges. This is because hyperedges can represent functional modules composed of various nodes, where the failure times of hyperedges bear significant practical implications. Moreover, activating a larger number of hyperedges may benefit more nodes' activation.

Based on the above analysis, we design a fitness function, which has three components for considering the joint influence of nodes and hyperedges.
The first component ${\sigma _0}(S)$ indicates the size of the seed set, the second component ${\sigma _1}(S)$ denotes the expected number of failed hyperedges associated with the seed set, and the third component ${\sigma _2}(S)$ represents the expected number of the failed first-layer neighboring nodes of the seed set. It can be mathematically defined as follows:

\begin{equation}
\begin{aligned}
W(S) & = \sigma_0(S) + \sigma_1(S) + \sigma_2(S) \\
& = \left| S \right| + \sum_{{e_\gamma} \in E_S^{(1)}} \left(1 - \prod_{i \in S \atop H_{i{e_\gamma}} = 1} (1 - t_{i{e_\gamma}}) \right) \\ 
&+ \sum_{\mu \in N_S^{(1)}/S} \left(1 - \prod_{e_\gamma \in E_S^{(1)} \atop H_{\mu{e_\gamma}} = 1} (1 - p_{e_\gamma}s_{{e_\gamma}\mu}) \right)
\end{aligned}
\label{fitness}
\end{equation}

Where $S$ represents a seed set, $E_{S}^{(1)}$ denotes the set comprising all the hyperedges associated with $S$. $N_{S}^{(1)}$ represents the set of neighboring nodes to all the nodes in $S$. ${p_{{e_\gamma }}} = (1 - \prod\limits_{i \in S\hfill\atop
{H_{i{e_\gamma }}} = 1\hfill} {1 - {s_{i{e_\gamma }}}} )$ represents the probability that the hyperedge $e_\gamma$ fails. 

\begin{figure}
	\centering
		\includegraphics[scale=.38]{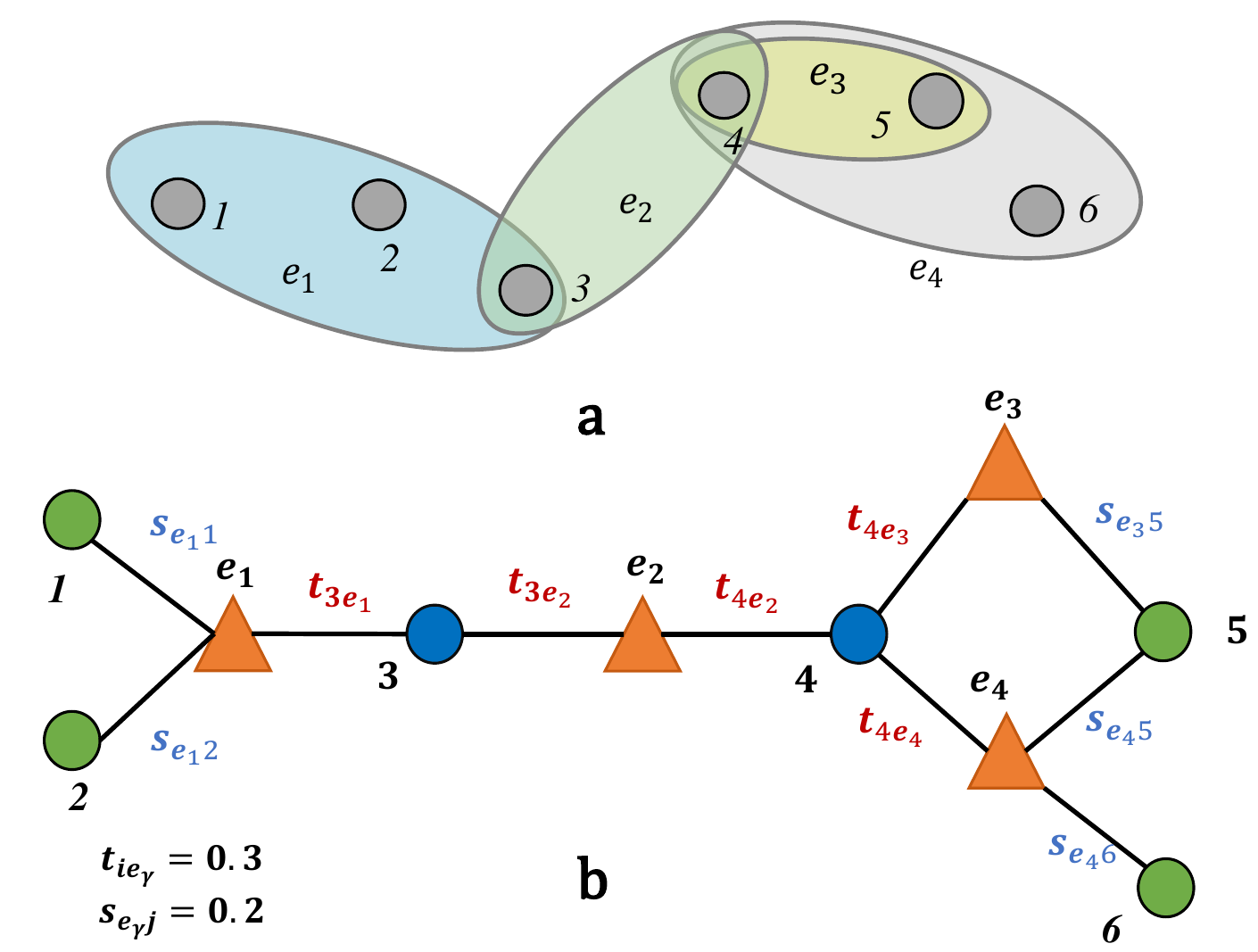}
	\caption{The illustrates of the fitness function. (a) shows a hypergraph with $6$ nodes and $4$ hyperedges. (b) represents an alternative representation of the hypergraph, where green circles denote nodes, orange triangles denote hyperedges, and the links represent the association relationship between nodes and hyperedges. 
 }
\label{FIG:2}
\end{figure}

We take an example illustrated in Fig. \ref{FIG:2} to explain how the fitness function works. At the beginning, the propagation probabilities, i.e., $t_{i e_\gamma}$ and $s_{e_\gamma i}$, are set to $0.3$ and $0.2$, respectively, and the seed nodes are given as $S=\{3,4\}$. Then, we can obtain $E_{S}^{(1)}=\{e_1, e_2, e_3, e_5\}$ and $N_{S}^{(1)}=\{1, 2, 5, 6\}$. Therefore, the first component ${\sigma _0}(S)$ is $2$, the second component ${\sigma _1}(S)$ is $1.41$ (i.e., 0.3+0.51+0.3+0.3), and the third component ${\sigma _2}(S)$ is $0.2964$ (i.e., 0.06+0.06+0.1164+0.06). Therefore, the fitness value is $W(S)=2+1.41+0.2964=3.7064$.

\subsection{Crossover and Mutation Operators}
\begin{figure}
	\centering
		\includegraphics[scale=.5]{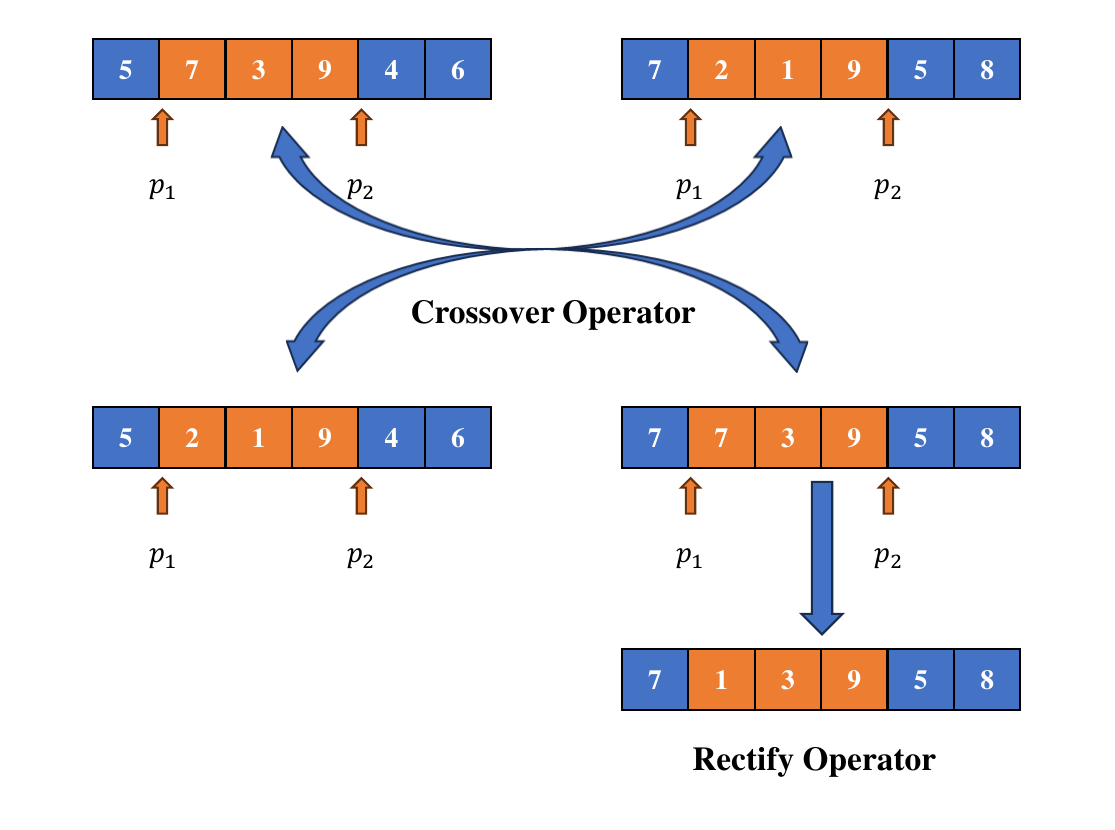}
	\caption{The illustrates of the proposed crossover operator. The crossover operation uses two-point crossover, and if there are identical nodes in the crossed individuals, one of them is replaced by a random node to ensure that there are no two identical nodes in an individual. 
}
\label{FIG:Crossover}
\end{figure}

We employ a two-point crossover operator \cite{Wang2021900} in this study. 
As illustrated in Fig. \ref{FIG:Crossover}, two points, $p_1$ and $p_2$, are randomly generated, where $p_1 < p_2$. For the two parent individuals, the segments between $p_1$ and $p_2$ are exchanged to create two new offspring. Suppose there are duplicate nodes selected by an individual after crossover. In that case, they are randomly replaced by nodes that have not been included in the individual, as the example in Fig. \ref{FIG:Crossover}. 

The clustering effect of influential nodes often hampers information propagation. To address this issue, we define a comprehensive metric for the mutation operator that simultaneously considers both the hypergraph collective influence and their overlapping influence. Firstly, the overlapping influence (OI) of nodes can be defined as the non-overlapping ratio of a node's first-layer neighbors to all the neighbors in an individual except the mutation point. This can be expressed as:

\begin{equation}
OI_i = \frac{{\left| {NS(i) - NS(T)} \right|}}{{\left| {NS(i)} \right|}}
\label{OF}
\end{equation}

where $NS(i)$ is the set of direct neighbor nodes of node $i$, $NS(T)$ is the set of all direct neighbors in an individual except the mutation point. Then a comprehensive metric (CM) considering both the hypergraph collective influence and their overlapping influence is defined as follows:

\begin{equation}
CM_i = OI_i \times HC{I_i} - ICM
\label{CM}
\end{equation}

\begin{figure}
	\centering
		\includegraphics[scale=.55]{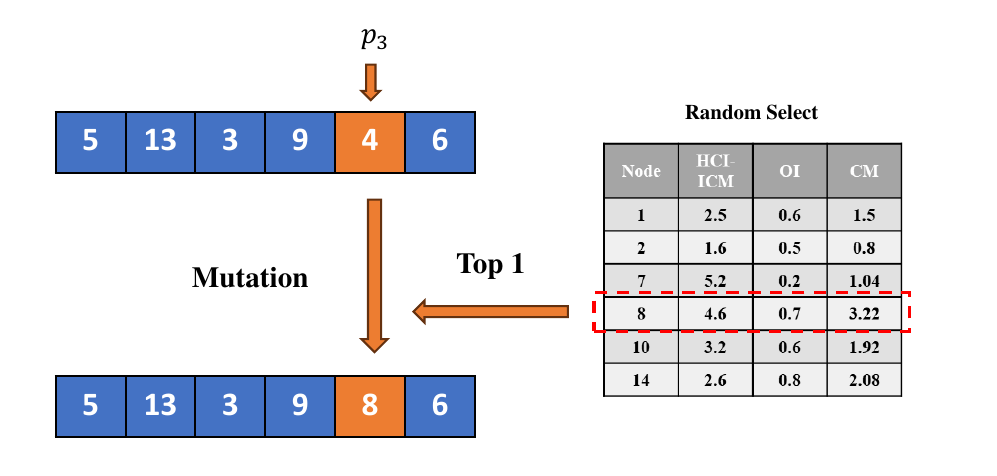}
	\caption{The illustrates of mutation operator. In the mutation operator, we use multi-point mutation, and each element in the individual has a certain probability of mutation. In the process of mutation, we randomly select $k$ nodes from the candidate set and consider factors such as the collective influence of nodes and the overlap influence of nodes to establish a comprehensive metric and select the node with the largest comprehensive metric as the node after mutation.
}
\label{FIG:Mutation}
\end{figure}

In the mutation process, we employ the multi-point mutation with the probability of $MUTPB$, as illustrated in Fig. \ref{FIG:Mutation}. Each gene (i.e., node) in the individual has a certain probability of being mutated, which is set to $1/len(individual)$ in this study. For each mutation point, we select $k$ nodes from the candidate set including all the other nodes except the nodes in the individual, calculate their comprehensive metrics, and replace the mutation point with the node having the highest comprehensive metric. This operation facilitates the evolution of the population towards better solutions. 

\section{Experiment Design}

\subsection{Data Sets}
In the experiments, we tested the effectiveness of the proposed method on both synthetic and real-world hypergraphs to ensure comprehensive evaluations. These experiments guarantee a thorough examination of the proposed method's effectiveness.

In the simulations of the synthetic hypergraph experiments, a configuration method \cite{Xie2023} was employed to generate ER hypergraphs, SF hypergraphs, and K-UF hypergraphs. The details of hypergraph generation are provided in Table \ref{TB2}. To mitigate the influence of peripheral nodes, all the experiments were conducted on the largest connected component of the hypergraphs. The results obtained from the algorithms with stochastic properties, including the G-CIIM algorithm, G-CI algorithm, GA algorithm, and RD algorithm, have been averaged over the 30 runs to ensure the reliability and reproducibility of the findings. We also validated the algorithm on three real-world hypergraph datasets, including the Contact-high-school dataset \cite{Chodrow2021, Mastrandrea2015}, the House-committees \cite{Chodrow2021} dataset, and the Restaurant-rev dataset \cite{Amburg2022145}. The specific characteristics of these datasets and the settings for the propagation probabilities are summarized in Table \ref{TB3}.

\begin{table}
\centering
\caption{The properties of synthetic hypergraphs.}
\begin{tabular}{cccc}
\hline
Network&  N& M& Feature\\
\hline
SF& 2000& 500& Power-law Exponents $\lambda=-2$\\
K-UF& 2000& 2000& Cardinality of hyperedges $m=5$\\
\hline
\end{tabular}
\label{TB2}
\end{table}

\begin{table}
\centering
\caption{The properties of real-world hypergraphs.}
\begin{tabular}{ccccccc}
\hline
Network&  $N$& $M$& $\left\langle k \right\rangle$& $\left\langle m \right\rangle$& $t_{i e_\gamma}$& $s_{e_\gamma i}$\\
\hline
House-committees& 1,290& 341& 9.2& 34.8& 0.05& 0.05\\
Contact-high-school& 327& 7,818& 55.6& 2.3& 0.1& 0.1\\
Restaurant-rev& 565& 601& 8.1& 7.7& 0.1& 0.1\\
\hline
\end{tabular}
\label{TB3}
\end{table}

\subsection{Baseline Methods}
In simulations, we compared the G-CIIM method with eight baseline algorithms, including the G-CI algorithm (using original random mutation operation), the standard GA algorithm, High-Hyper-Degree algorithm (HHD), First-order Hypergraph Collective Influence algorithm (HCI1), Second-order Hypergraph Collective Influence algorithm (HCI2), Neighbor Priority algorithm (NP) \cite{Zhang2024}, PageRank algorithm (PR) \cite{Zhang2024}, and Random algorithm (RD). The propagation scope of all experiments was based on the results of 10,000 Monte Carlo simulations.

\subsection{Parameter Settings}
All the experiments were conducted on a PC equipped with a 5.6 GHz Intel i7 CPU and 32 GB RAM. The entire codebase was implemented in Python and utilized the DEAP framework. During the initialization process of the G-CIIM method, we leverage the collective influence of the hypergraph to initialize the population, to obtain superior initial solutions. To prevent the algorithm from getting trapped in local optima, we set a relatively high mutation 
rate of 0.5. This is to enhance the algorithm's exploration ability to escape local optima and discover potentially better solutions. The parameter settings of the G-CIIM algorithm are described in Table \ref{TB1}.

\begin{table}
\centering
\caption{The parameter seting of G-CIIM algorithm.}
\begin{tabular}{ccc}
\hline
Parameter& Meaning& Value\\
\hline
Popnum& The size of initial population& 512\\
CXPB& Probability of performing a crossover operation& 0.5\\
MUTPB& Probability of performing a mutation operation& 0.5\\
Maxgen& The maximum number of generations& 100\\
Tournsize& Number of individuals selected for each tournament& 5\\
\hline
\end{tabular}
\label{TB1}
\end{table}

\section{Result Analysis and Discussions}


\subsection{Experiments on Synthetic Hypergraphs}

First, we conducted experimental simulations on Erdős-Rényi (ER) hypergraph, scale-free (SF) hypergraph, and k-uniform (K-UF) hypergraphs, respectively. In the process of IC Model propagation, an overly large propagation probability leads all the algorithms in the experiments to select a small number of seed nodes to achieve a significant propagation scope. However, an excessively small propagation probability might lead all the algorithms to choose diverse seed nodes yet remain confined to a limited propagation scope. To distinguish the performance of different algorithms, we set the propagation probability $s_{e_\gamma i}$ and $t_{i e_\gamma}$ to 0.15.

\begin{figure*}
	\centering
		\includegraphics[scale=.32]{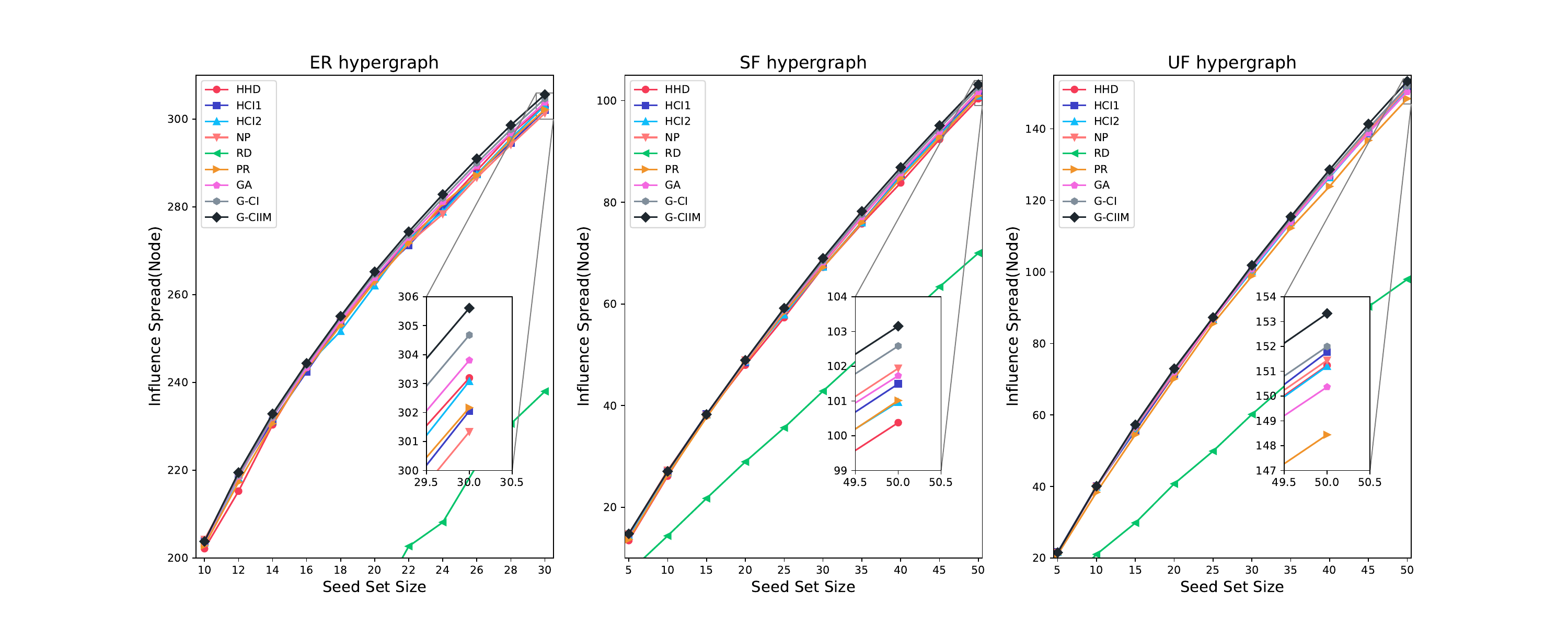}
	\caption{The performance of G-CIIM algorithm and other comparison algorithms on synthetic hypergraphs. (a-c) represent the performance of the algorithms on ER, SF, and UF hypergraphs. The horizontal axis represents the number of seed nodes, the vertical axis represents the spread influence (the number of activated nodes).
}
\label{FIG:sythetic-res}
\end{figure*}

The experimental results of the G-CIIM method and other baseline algorithms on ER, SF, and K-UF hypergraphs are illustrated in Fig. \ref{FIG:sythetic-res}. It can be observed that the G-CIIM method consistently outperforms the baseline methods, and its superiority becomes more evident as the number of seed nodes increases. This is due mainly to the increasing complexity of the optimization problem with a larger number of seed nodes, where the strong search capability of G-CIIM effectively overcomes this challenge, unlike centrality-based algorithms that struggle with such complexities. Moreover, it is found that G-CIIM outperforms both the G-CI and standard GA methods. This validates the effectiveness of the initialization and mutation operators. Apart from the RD algorithm, the performance differences among the other baseline methods are relatively minor. This may be due to the inherently more complex higher-order relationships presented in hypergraphs, posing greater optimization challenges while requiring further analysis and research on the dynamics of hypergraph-independent cascade models.

\subsection{Experiments on Real-world Hypergraphs}

To further validate the practical utility of the G-CIIM method, we conducted experimental simulations on three real-world hypergraph datasets, including the Contact-high-school dataset \cite{Chodrow2021, Mastrandrea2015}, the House-committees \cite{Chodrow2021} dataset, and the Restaurant-rev dataset \cite{Amburg2022145}. These datasets offer diverse representations of hypergraph structures, enabling us to comprehensively assess the performance of G-CIIM in real-world scenarios.

\begin{figure*}
	\centering
		\includegraphics[scale=.32]{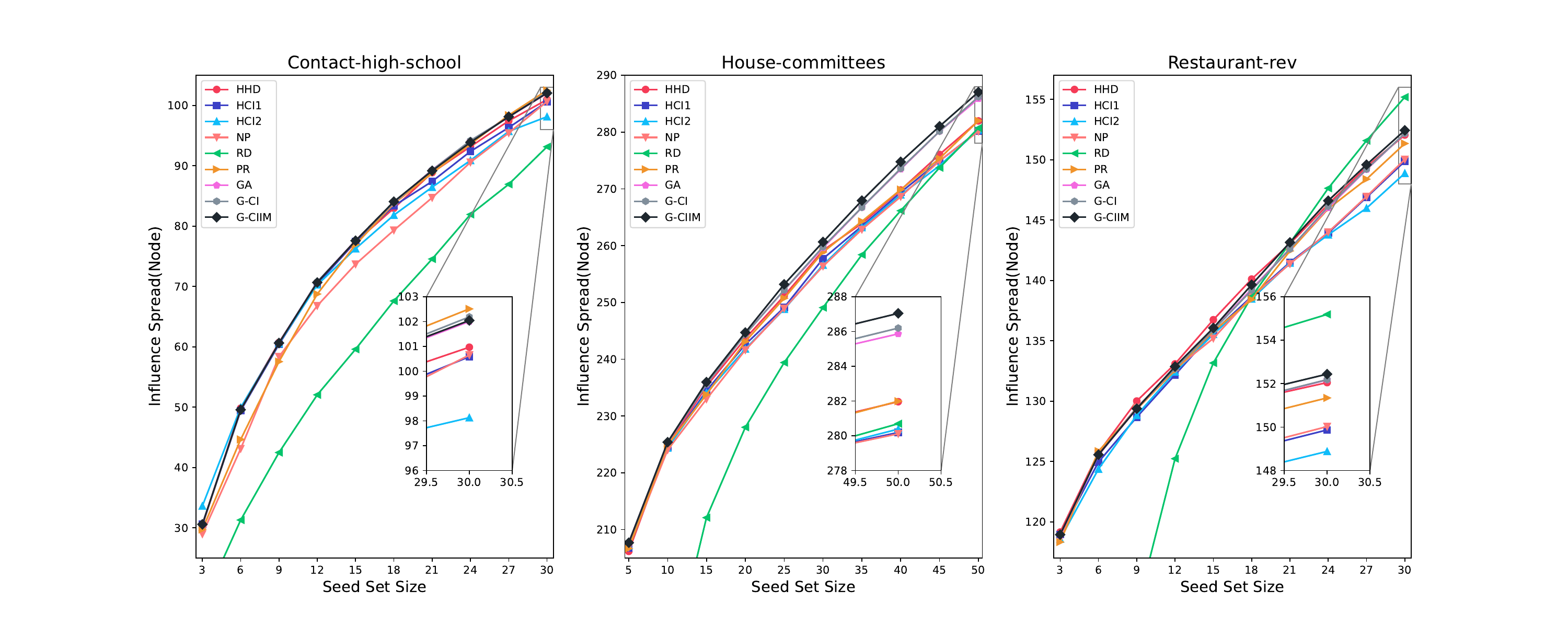}
	\caption{The performance of G-CIIM and baseline methods on real-world hypergraphs. (a-c) represent the performance of these methods on House-committees, Contact-high-school, and Restaurant-rev hypergraphs. The horizontal axis represents the number of seed nodes, and the vertical axis represents the spread influence (the number of activated nodes).
}
\label{FIG:real-res}
\end{figure*}

\begin{figure*}
	\centering
		\includegraphics[scale=.32]{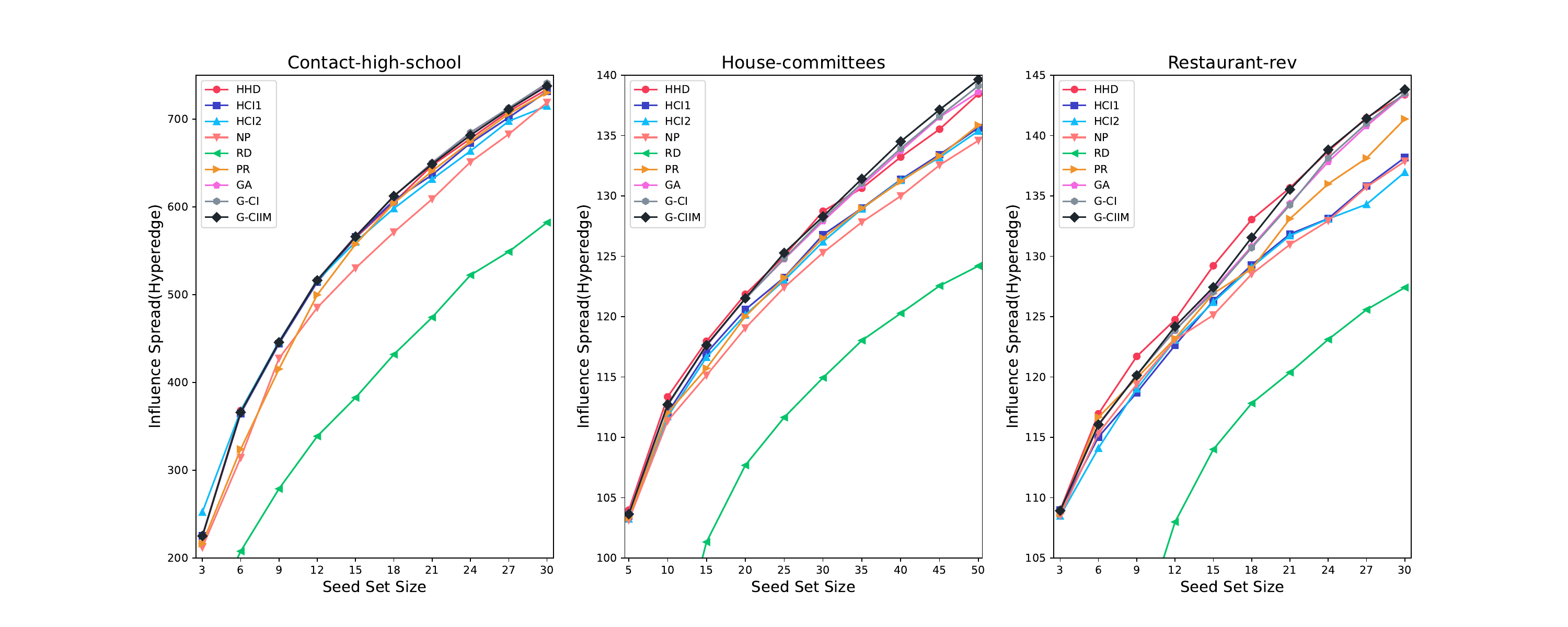}
	\caption{The performance of G-CIIM and other baseline methods on real-world hypergraphs. (a-c) represent the performance of these methods on House-committees, Contact-high-school, and Restaurant-rev hypergraphs. The horizontal axis represents the number of seed nodes, and the vertical axis represents the spread influence (the number of activated hyperedges).
}
\label{FIG:real-res-edge}
\end{figure*}

The experimental results of the G-CIIM method compared with other benchmark algorithms on real-world hypergraphs are depicted in Fig. \ref{FIG:real-res}. The advantage of the G-CIIM algorithm becomes increasingly pronounced as the number of seed nodes increases, indicating its efficacy in solving larger search spaces. Additionally, it can be observed that on the House-committees and Restaurant-rev datasets, the RD algorithm exhibits a sharp increase in the spread range with an increasing number of seed nodes. Remarkably, on the Restaurant-rev dataset, when 30 seed nodes are selected, the RD algorithm surpasses all the other algorithms. This phenomenon may be attributed to the clustering effect of influential nodes, which hinders information dissemination to a broader audience. Then, we plotted the line graph of the corresponding number of failed hyperedges for different numbers of seed nodes for the G-CIIM algorithm and other benchmark algorithms, as shown in Fig. \ref{FIG:real-res-edge}. The G-CIIM method outperforms other algorithms, particularly in Fig. \ref{FIG:real-res}c, where although the RD algorithm activates more nodes than the G-CIIM algorithm, it lags significantly behind in the number of failed hyperedges. This suggests that G-CIIM can identify nodes that exert the most significant influence on both hyperedges and nodes simultaneously.

\subsection{Analysis of Ablation Experiments}

\begin{figure*}
	\centering
		\includegraphics[scale=.3]{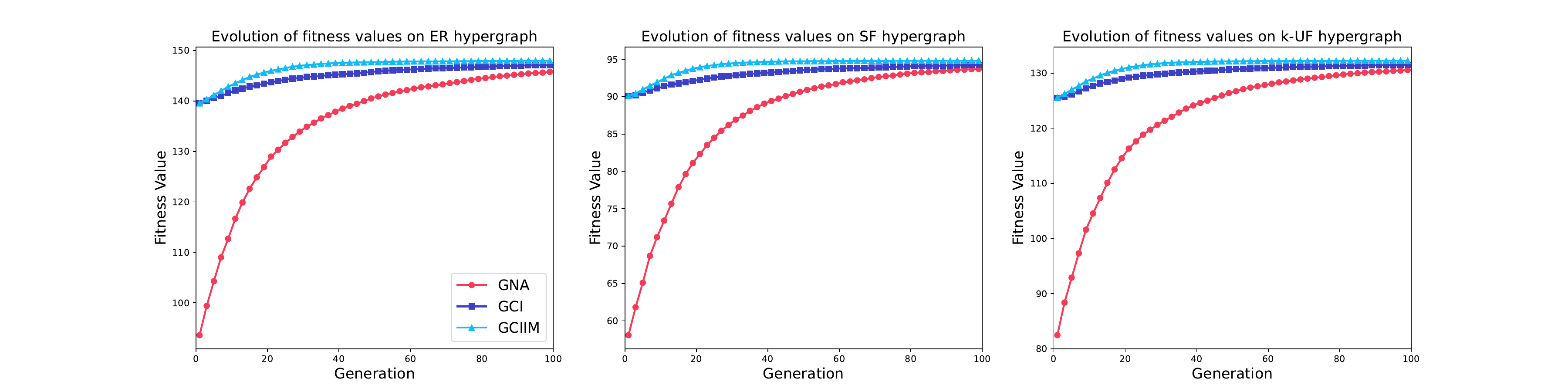}
	\caption{The illustrates of the changes of fitness values with generations for G-CIIM, G-CI and GA algorithms. (a-c) represent the changes of fitness value on ER hypergraph, SF hypergraph and K-UF hypergraph. 
}
\label{FIG:analysis}
\end{figure*}

To validate the effectiveness of the initialization and mutation operators, we plotted the evolutionary trajectories of the fitness values over generations for the GA, G-CI, and G-CIIM algorithms on ER, SF, and K-UF hypergraphs, as shown in Fig. \ref{FIG:analysis}. We set the number of iterations for all algorithms to 100. From Fig. \ref{FIG:analysis}, it is observed that the G-CI and G-CIIM methods, initialized with hypergraph collective influence, exhibited superior population quality and higher fitness values compared to the standard GA method at the beginning of the iterations. The initialization operator enables the algorithms to evolve high-quality solutions with fewer iterations. Furthermore, to compare G-CI and G-CIIM by analyzing their evolutionary processes, it is found that the mutation operator accelerates the evolution of the population, guiding it toward better solutions. At 100 generations, it is found that the fitness value of G-CIIM exceeded that of G-CI, which in turn exceeded that of GA. This demonstrates the effectiveness of the initialization and mutation operators in enabling the G-CIIM method to evolve superior solutions.

Experimental simulations have been conducted on ER hypergraphs to validate whether the collective influence-based method can select more influential nodes than degree-based method. The experimental results, presented in Appendix B, demonstrate the effectiveness of our approach. In the final analysis, we conducted Wilcoxon rank-sum tests (the the signiﬁcance level is set to 0.05) to compare the G-CIIM method with other baseline methods, with statistical results presented in Tab \ref{TBST}. In the synthetic hypergraphs, it is evident from the table that the G-CIIM method outperforms other baseline methods significantly in all the comparisons. In the real-world hypergraphs, the G-CIIM algorithm also demonstrates a significant superiority over alternative algorithms in almost all cases. This further indicates the effectiveness of the proposed method.

\begin{table}[]
\centering
\caption{Statistical results of G-CIIM algorithm and other comparison algorithms on synthetic hypergraphs and real-world hypergraphs.}
\label{TBST}
\begin{tabular}{llll}
\hline
Hypergraph                           & Algorithm & Average Influence & Std.  \\ \hline
\multirow{9}{*}{ER (k=30)}                
                                     & G-CI      & 304.67 
                                     +         & 0.36  \\  
                                     & GA        & 303.80 +         & 0.33  \\  
                                     & HHD       & 302.50 +         & 0.36  \\  
                                     & HCI1      & 302.60 +          & 0.36  \\  
                                     & HCI2      & 302.82 +          & 0.34  \\  
                                     & NP        & 301.01 +         & 0.33  \\ 
                                     & RD        & 235.44 +          & 46.25 \\ 
                                     & PR        & 301.82 +         & 0.23  \\ 
                                       & G-CIIM    & 305.61            & 0.3   \\ \hline
\multirow{9}{*}{SF (k=50)}                 
                                     & G-CI      & 102.58 +         & 0.08  \\ 
                                     & GA        & 101.72 +         & 0.22  \\ 
                                     & HHD       & 100.43 +         & 0.01  \\ 
                                     & HCI1      & 101.56 +         & 0.01  \\ 
                                     & HCI2      & 101.04 +         & 0.01  \\ 
                                     & NP        & 101.94 +          & 0.01  \\ 
                                     & RD        & 70.28 +           & 3.32  \\ 
                                     & PR        & 101.02 +         & 0.01  \\ 
                                      & G-CIIM    & 103.15            & 0.04  \\ \hline
\multirow{9}{*}{UF (k=50)}               
                                     & G-CI      & 151.99 +         & 0.22  \\ 
                                     & GA        & 150.36 +         & 0.56  \\ 
                                     & HHD       & 151.56 +         & 0.04  \\ 
                                     & HCI1      & 151.57 +         & 0.05  \\ 
                                     & HCI2      & 151.40 +         & 0.036 \\ 
                                     & NP        & 151.44 +         & 0.04  \\ 
                                     & RD        & 99.57 +           & 12.08 \\ 
                                     & PR        & 148.28 +         & 0.05  \\ 
                                        & G-CIIM    & 153.32         & 0.06  \\ \hline
\multirow{9}{*}{Contact-high-school (k=30)} 
                                     & G-CI      & 102.19 =          & 0.27  \\ 
                                     & GA        & 102.00 =         & 0.25  \\ 
                                     & HHD       & 100.73 +         & 0.01  \\ 
                                     & HCI1      & 100.72 +         & 0.01  \\ 
                                     & HCI2      & 98.1 +           & 0.02  \\ 
                                     & NP        & 100.84 +    & 0.02  \\ 
                                     & RD        & 93.48 +          & 2.51  \\ 
                                     & PR        & 102.51 -         & 0.02  \\ 
                                      & G-CIIM    & 102.04            & 0.33  \\ \hline
\multirow{9}{*}{House-committees (k=50)}    
                                     & G-CI      & 286.19 +         & 0.26  \\ 
                                     & GA        & 285.86 +         & 0.34  \\ 
                                     & HHD       & 281.95 +         & 0.08  \\ 
                                     & HCI1      & 280.62 +         & 0.04  \\ 
                                     & HCI2      & 280.01 +         & 0.06  \\ 
                                     & NP        & 280.3 +          & 0.08  \\ 
                                     & RD        & 280.72 +         & 2.1   \\ 
                                     & PR        & 281.19 +         & 0.04  \\ 
                                     & G-CIIM    & 287.04            & 0.24  \\ \hline
\multirow{9}{*}{Restaurant-rev (k=30)}     
                                     & G-CI      & 152.18 +         & 0.07  \\ 
                                     & GA        & 152.17 +         & 0.08  \\ 
                                     & HHD       & 151.97 +         & 0.02  \\ 
                                     & HCI1      & 149.86 +         & 0.02  \\ 
                                     & HCI2      & 149.03 +         & 0.02  \\ 
                                     & NP        & 149.99 +         & 0.02  \\ 
                                     & RD        & 154.98 -         & 0.67  \\ 
                                     & PR        & 151.42 +         & 0.02  \\ 
                                      & G-CIIM    & 152.44            & 0.09  \\ \hline
\end{tabular}
\end{table}

\section{Conclusions}

This paper aims to address the IM problem in hypergraphs by utilizing GA. The goal has been successfully achieved by proposing the G-CIIM method. In G-CIIM, a new IC model is first proposed, which offers the model framework to G-CIIM with a more comprehensive understanding of influence propagation dynamics on hypergraph structures. Then, GA is introduced to identify the most influential nodes. In G-CIIM, the hypergraph collective influence is used to enhance the quality of initial solutions. To effectively evaluate the quality of individuals, the designed fitness function simultaneously considers the impacts of nodes and hyperedges, to identify node sets with optimal influence on both nodes and hyperedges. By integrating factors such as the hypergraph collective influence and the overlapping influence of nodes, a mutation operator is designed, to guide the population towards higher-quality solutions. 

In the experiments, G-CIIM demonstrates its superiority over the baseline algorithms on synthetic and real-world hypergraphs. More importantly, its superiority becomes increasingly apparent as the number of seed nodes increases. However, in real-world hypergraphs, we also observe that nodes with significant influence typically exhibit clustering effects, hindering the broader dissemination of information. 
In addition, due to the higher-order interactions between nodes and hyperedges, the dynamic behaviors of the IC model may be different. It would be interesting to investigate the dynamics of this model in the future.


\appendix

\section*{A. The Derivation of Hypergraph Collective Influence}

In order to improve the efficiency of the genetic algorithm, this paper uses hypergraph collective influence to initialize the population and design the mutation operator. The hypergraph collective influence is a node centrality method to consider the local structure of hypergraphs. It is obtained by analyzing the stability of equilibrium solutions of self-satisfying equations using the Message Passing theory framework \cite{Zhang2024}. The detail of the derivation process is expressed below.

First, to weaken the strong correlation between nodes, the conditional probability self-satisfying equations satisfying the independent cascade rule of hypergraphs are established by using Cavity method. It can be expressed as:

\begin{equation}
\left\{ {\begin{array}{*{20}{c}}
{v_{i \to {e_\gamma }}^{t + 1} = {n_i} + (1 - {n_i})[1 - \prod\limits_{{e_\beta } \in \partial i/{e_\gamma }} {(1 - {s_{{e_\beta }i}}v_{{e_\beta } \to i}^t)} ]}\\
{v_{{e_\gamma } \to i}^{t + 1} = 1 - \prod\limits_{p \in {e_\gamma }/i} {(1 - {t_{p{e_\gamma }}}v_{p \to {e_\gamma }}^t)} }
\end{array}} \right.
\label{MP}
\end{equation}

$v_{i \to e_\gamma}$ represents the failure probability of node $i$ assuming hyperedge $e_\gamma$ is removed, $v_{e_\gamma \to i}$ represents the failure probability of hyperedge $e_\gamma$ assuming node $i$ is removed. To simplify Eq.(\ref{MP}), we set ${V_ \to } = {\{ {\vec v_1},{\vec v_2}\} ^T}$, ${n_ \to } = [{\vec n_1},0]_{1 \times 2d} = {(...,\underbrace {{n_i},...,{n_i}}_{{k_i}},...,\underbrace {0,...,0}_d)^T}$, where ${\vec v_1} = {[{v_{i \to {e_\gamma }}}]_{1 \times d}}$, ${\vec v_2} = {[{v_{{e_\gamma } \to i}}]_{1 \times d}}$, $d = \sum\limits_{i = 1}^N {{k_i}}$ denotes the sum of hyperdegree. Therefore, Eq.(\ref{MP}) can be expressed as:

\begin{equation}
V_ \to ^{t + 1} = {n_ \to } + F(V_ \to ^t) \Leftrightarrow \left\{ {\begin{array}{*{20}{c}}
{v_1^{t + 1} = {n_1} + {f_1}(v_2^t)}\\
{v_2^{t + 1} = {f_2}(v_1^t)}
\end{array}} \right.
\label{MP2}
\end{equation}

Due to the complexity of Eq.(\ref{MP2}), we linearize it to analysis conveniently. The Jacobian matrix can be expressed as:

\begin{equation}
{\cal J} = {\left( {\begin{array}{*{20}{c}}
{\frac{{\partial {f_1}}}{{\partial {{\vec v}_1}}}}&{\frac{{\partial {f_1}}}{{\partial {{\vec v}_2}}}}\\
{\frac{{\partial {f_2}}}{{\partial {{\vec v}_1}}}}&{\frac{{\partial {f_2}}}{{\partial {{\vec v}_2}}}}
\end{array}} \right)_{2d \times 2d}}
\label{JB}
\end{equation}

Therefore the above nonlinear self-satisfying Eq. (\ref{MP2}) can be approximated by the following linear self-satisfying equations:

\begin{equation}
V_ \to ^{t + 1} = {n_ \to } + {\cal J}V_ \to ^t
\label{LQ}
\end{equation}

The final state of the Eq.(\ref{LQ}) can be expressed as:

\begin{equation}
V_ \to ^* = {n_ \to } + {\cal J}V_ \to ^*
\label{SQ}
\end{equation}

Then let's derive the concrete form of the Jacobian matrix. First, the partial derivative for $f_1$ is:

\begin{equation}
\frac{{\partial {v_{i \to {e_\gamma }}}}}{{\partial {v_{j \to {e_\beta }}}}} = 0
\label{f1-1}
\end{equation}

{\small
\begin{equation}
\frac{{\partial {v_{i \to {e_\gamma }}}}}{{\partial {v_{{e_\beta } \to j}}}} = \left\{ {\begin{array}{*{20}{c}}
{{s_{{e_\beta }j}}(1 - {n_i})\prod\limits_{{e_\mu } \in \partial i/{e_\gamma },{e_\beta }} {(1 - {s_{{e_\mu }i}}{v_{{e_\mu } \to i}})} }&{i = j,{e_\beta } \ne {e_\gamma }}\\
0&{otherwise}
\end{array}} \right.
\label{f1-2}
\end{equation}
}

So we can get the non-backtracking matrix when ${V_ \to } = {\{ {\vec 0},{\vec 0}\} ^T}$:

\begin{equation}
{\left. {{{\cal A}_{{e_\beta } \to j,i \to {e_\gamma }}} = \frac{{\partial {v_{i \to {e_\gamma }}}}}{{\partial {v_{{e_\beta } \to j}}}}} \right|_{(0,0)}} = \left\{ {\begin{array}{*{20}{c}}
{{s_{{e_\beta }j}}(1 - {n_i})}&{i = j,{e_\beta } \ne {e_\gamma }}\\
0&{otherwise}
\end{array}} \right.
\label{f1-3}
\end{equation}

Then let's compute the partial derivative of $f_2$:

\begin{equation}
\frac{{\partial {v_{{e_\gamma } \to i}}}}{{\partial {v_{{e_\beta } \to j}}}} = 0
\label{f2-1}
\end{equation}

\begin{equation}
\frac{{\partial {v_{{e_\gamma } \to i}}}}{{\partial {v_{j \to {e_\beta }}}}} = \left\{ {\begin{array}{*{20}{c}}
{{t_{j{e_\beta }}}\prod\limits_{{e_\mu } \in \partial i/{e_\gamma },{e_\beta }} {(1 - {t_{i{e_\mu }}}{v_{{e_\mu } \to i}})} }&{i \ne j,{e_\beta } = {e_\gamma }}\\
0&{otherwise}
\end{array}} \right.
\label{f2-2}
\end{equation}

Same as above, we can get another non-backtracking matrix through Eq.(\ref{f2-2}):

\begin{equation}
{\left. {{{\cal B}_{j \to {e_\beta },{e_\gamma } \to ei}} = \frac{{\partial {v_{{e_\gamma } \to i}}}}{{\partial {v_{j \to {e_\beta }}}}}} \right|_{(0,0)}} = \left\{ {\begin{array}{*{20}{c}}
{{t_{j{e_\beta }}}}&{i \ne j,{e_\beta } = {e_\gamma }}\\
0&{{\rm{otherwise}}}
\end{array}} \right.
\label{f2-3}
\end{equation}

Therefore the specific form of the Jacobian matrix can be described as:

\begin{equation}
{\cal J} = {\left[ {\begin{array}{*{20}{c}}
0&{\cal A}\\
{\cal B}&0
\end{array}} \right]_{2d \times 2d}}
\label{JBD}
\end{equation}

where ${\cal A} = {\{ {{\cal A}_{{e_\beta } \to j,i \to {e_\gamma }}}\} _{s \times s}}$, ${\cal B} = {\{ {{\cal B}_{j \to {e_\beta },{e_\gamma } \to i}}\} _{s \times s}}$. In order to further derivation, we extend it to a 4-dimensional tensor:

\begin{equation}
\left\{ {\begin{array}{*{20}{c}}
{{{\cal A}_{{e_\beta }ji{e_\gamma }}} = (1 - {n_j}){H_{j{e_\beta }}}{H_{i{e_\gamma }}}{\delta _{ij}}(1 - {\delta _{{e_\beta }{e_\gamma }}}){s_{{e_\beta }j}}}\\
{{{\cal B}_{j{e_\beta }{e_\gamma }i}} = {H_{j{e_\beta }}}{H_{i{e_\gamma }}}{\delta _{{e_\beta }{e_\gamma }}}(1 - {\delta _{ij}}){t_{j{e_\beta }}}}
\end{array}} \right.
\label{Tensor}
\end{equation}

For $t=1$, we set $V_ \to ^0 = {n_ \to }$. Therefore we can get $V_ \to ^1 = {n_ \to } + {\cal J}{n_ \to }$:

\begin{equation}
{\left[ {\begin{array}{*{20}{c}}
{{v_1}}\\
{{v_2}}
\end{array}} \right]^1} = \left[ {\begin{array}{*{20}{c}}
{{{\bf{n}}_1}}\\
{\bf{0}}
\end{array}} \right] + \left[ {\begin{array}{*{20}{c}}
0&{\cal A}\\
{\cal B}&0
\end{array}} \right]\left[ {\begin{array}{*{20}{c}}
{{{\bf{n}}_1}}\\
{\bf{0}}
\end{array}} \right] = \left[ {\begin{array}{*{20}{c}}
{{{\bf{n}}_1}}\\
{{\cal B}{{\bf{n}}_1}}
\end{array}} \right]
\label{T1}
\end{equation}

The specific form of each element in the vector can be expressed as:

\begin{equation}
\left\{ {\begin{array}{*{20}{l}}
{v_{i \to {e_\gamma }}^1 = {n_i}{H_{i{e_\gamma }}}}\\
{v_{{e_\gamma } \to i}^1 = {H_{i{e_\gamma }}}\sum\limits_j {{n_j}{H_{j{e_\gamma }}}(1 - {\delta _{ij}}){t_{j{e_\gamma }}}} }
\end{array}} \right.
\label{T1-1}
\end{equation}

Then we define $\left\| {{v_ \to }} \right\| = \sum\limits_{i{e_\gamma }} {({v_{i \to {e_\gamma }}} + {v_{{e_\gamma } \to i}})}$ to assess the range of propagation. Our goal is to select a certain number of seeds to maximize the range of propagation. According to Eq.(\ref{T1}), we can get that:

\begin{equation}
\begin{aligned}
\| {v_ \to } \| &= \sum_{i {e_\gamma}} {v_{i \to {e_\gamma}}} + \sum_{i {e_\gamma}} {v_{{e_\gamma} \to i}} \\
&= \sum_{i {e_\gamma}} {n_i}{H_{i{e_\gamma}}} + \sum_{i {e_\gamma}} {H_{i{e_\gamma}}} \sum_{j} {n_j}{H_{j{e_\gamma}}}(1 - {\delta _{ij}}){t_{j{e_\gamma}}} \\
&= \sum_{i} {n_i}{k_i} + \sum_{i} {n_i} \sum_{{e_\gamma} \in \partial i} \sum_{j \in \partial {e_\gamma}/i} {{t_{i{e_\gamma}}}} \\
&= \sum_{i} {n_i} \left( {{k_i} + \sum_{{e_\gamma} \in \partial i} \sum_{j \in \partial {e_\gamma}/i} {{t_{i{e_\gamma}}}} } \right)\\
&= \sum_{i} {n_i} \left( {{k_i} + \sum_{{e_\gamma} \in \partial i} {{t_{i{e_\gamma}}}({m_\gamma} - 1)} } \right)
\end{aligned}
\label{Sum}
\end{equation}

To sum up, we can obtain the 1-order hypergraph collective influence base on the independent cascade model(HCI-ICM):

\begin{equation}
HC{I_1}(i) - ICM = {k_i} + \sum\limits_{{e_\gamma } \in \partial i} {{t_{i{e_\gamma }}}({m_\gamma } - 1)}
\label{HCI1}
\end{equation}

Same as above, we can get 2-oder HCI-ICM and n-order HCI-ICM:

\begin{equation}
\begin{aligned}
    HCI_2(i) - ICM &= {k_i} + \sum\limits_{{e_\gamma } \in \partial i} {{t_{i{e_\gamma }}}({m_\gamma } - 1)} \\
    &+ \sum\limits_{{e_\gamma } \in \partial i} {{t_{i{e_\gamma }}}\sum\limits_{j \in {e_\gamma }/i} {(1 - {n_j}){s_{{e_\gamma }j}}({k_j} - 1)} }
\end{aligned}
\label{HCI2}
\end{equation}

\begin{equation}  
\begin{split}  
HC{I_n}(i) - ICM = {k_i} + \sum\limits_{L \in {A_n}} {\Re _L^n}  + \sum\limits_{L \in {B_n}} {\aleph _L^n} 
\end{split}  
\label{HCIn}   
\end{equation}

Here ${A_n} = \{ x \in {N^ + }|x mod\;2 = 0,x \le n\}$, ${B_n} = \{ x \in {N^ + }|x\;mod\;2 = 1,x \le n\}$, ${\Re_L^n}$ and ${\aleph_L^n}$ are represent as:

\begin{equation}
\begin{split}
\Re _L^n &= \sum\limits_{{e_{{\gamma _1}}} \in \partial {i_1}} {{t_{{i_1}{e_\gamma }_1}}\sum\limits_{{i_2} \in \partial {e_{{\gamma _1}}}/{i_1}} {(1 - {n_{{i_1}}}){s_{{e_\gamma }_1{i_2}}}} }  \times  \cdots \\
&\times \sum\limits_{{i_\ell } \in \partial {e_{{\gamma _{\ell  - 1}}}}/{i_{\ell  - 1}}} {(1 - {n_{{i_l}}}){s_{{e_\gamma }_{l - 1}{i_l}}}({k_{{i_l}}} - 1)}
\end{split}
\label{HCIn-1}
\end{equation}

\begin{equation}
\begin{split}
\aleph _L^n &= \sum\limits_{{e_{{\gamma _1}}} \in \partial {i_1}} {{t_{{i_1}{e_\gamma }_1}}\sum\limits_{{i_2} \in \partial {e_{{\gamma _1}}}/{i_1}} {(1 - {n_{{i_1}}}){s_{{e_\gamma }_1{i_2}}}} }  \times  \cdots \\
&\times \sum\limits_{{e_{{\gamma _l}}} \in \partial {i_\iota }/{e_{{\gamma _{l - 1}}}}} {{t_{{i_l}{e_{{\gamma _l}}}}}({m_{{\gamma _l}}} - 1)}
\end{split}
\label{HCIn-2}
\end{equation}

\section*{B. The performance of HHD vs HCI}

\begin{figure}
	\centering
		\includegraphics[scale=.3]{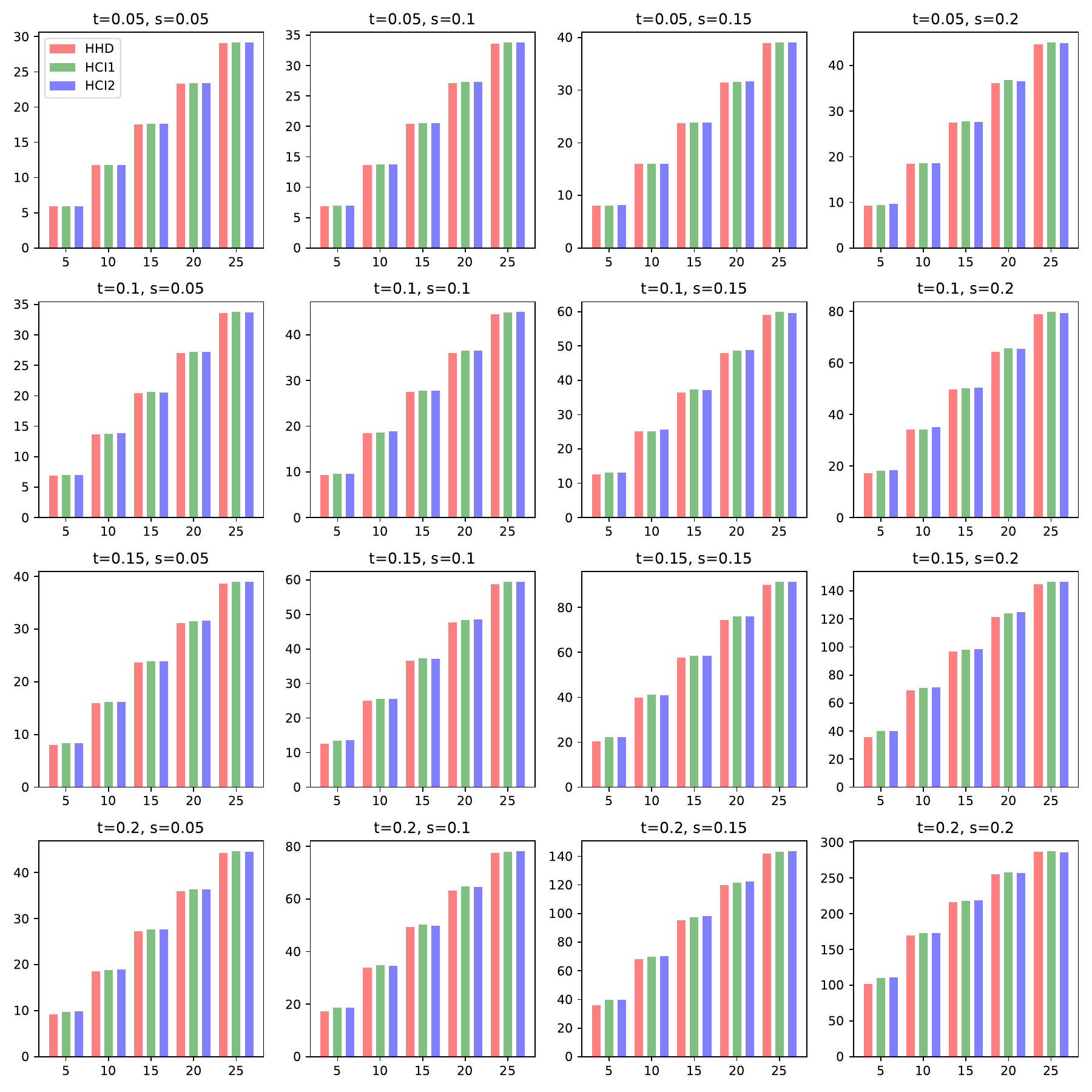}
	\caption{The performance of HHD and HCI algorithm on ER hypergraph under different parameter setting. The horizontal axis represents the number of seed, the vertical axis represents the number of activated nodes.}
\label{FIG:appendix}
\end{figure}

To demonstrate the superiority of the HCI algorithm in identifying the most influential nodes, we conducted a comparative analysis on ER hypergraphs. We set the propagation parameters $t_{i e_\gamma}$ and $s_{e_\gamma i}$ to range from $0.05$ to $0.2$, with an increment of $0.05$. For each parameter setting, we plotted bar charts to compare the activated nodes achieved by different algorithms across various seed numbers in Fig. \ref{FIG:appendix}. It is evident that the HCI1 algorithm exhibits superior performance in the most cases, effectively selecting nodes with greater influence potential. Therefore, we adopted the 1-order collective influence for initializing the population, which enhances the initial quality of the population.

%





\ifCLASSOPTIONcaptionsoff
  \newpage
\fi





\bibliographystyle{IEEEtran}
\bibliography{IEEEabrv,Bibliography}

\vfill


\end{document}